\documentclass[aps,pra,twocolumn,showpacs,showkeys,preprintnumbers,amsmath,amssymb]{revtex4}

\usepackage{graphicx}
\usepackage{dcolumn}
\usepackage{bm}

\DeclareMathOperator*{\im}{Im} \DeclareMathOperator*{\re}{Re}

\begin{document}

\preprint{APS/123-QED}

\title{Role of temperature effects in the phenomenon of
        ultraslow electromagnetic pulses\\ in Bose-Einstein
        condensates of alkali-metal atoms}

\author{Yurii Slyusarenko}
 \email{slusarenko@kipt.kharkov.ua}
\author{Andrii Sotnikov}%
\affiliation{%
Akhiezer Institute for Theoretical Physics, NSC KIPT,\\
1 Akademichna Street, Kharkiv 61108, Ukraine
}%

\date{\today}

\begin{abstract}
We study the temperature dependence of optical properties of dilute
gases of alkali-metal atoms in the state with Bose-Einstein
condensates. The description is constructed in the framework of the
microscopic approach that is based on the Green-functions formalism.
We find the expressions for the scalar Green functions describing a
linear response of a condensed gas to a weak external
electromagnetic field (laser). It is shown that these functions depend
on the temperature, other physical properties of a system, and on the
frequency detuning of a laser. We compare the
relative contributions of the condensate and non-condensate particles
in the system response. The influence of
the temperature effects is studied by the example of two- and
three-level systems. We show that in these cases, which are most
commonly realized in the present experiments, the group velocity and
the absorption rate of pulses practically do not depend on the gas
temperature in the region from the absolute zero to the critical
temperature. We discuss also the cases when the temperature effects
can play a significant role in the phenomenon of slowing of
electromagnetic pulses in a gas of alkali-metal atoms with
Bose-Einstein condensates.
\end{abstract}


\maketitle

\section{Introduction}

Bose-Einstein condensate (BEC) is one of the most impressive
examples when the matter demonstrates its quantum nature on the
macroscopic level. Now this system is interesting also due the
possibility of observing the electromagnetic pulses propagating with
extremely slow group velocities in it~\cite{Hau1999N}.

Up to now, in the theoretical investigations
pretending to describe the mentioned phenomenon in a BEC (see
Refs.~\cite{Hau2004PRA,Sly2008PRA}) the authors assumed that the
temperature of a gas is small in comparison to the critical
temperature. In other words, this assumption corresponds to the
consideration of the zero-temperature limit. But in real systems, the
temperatures can be of the same order as the critical temperature.
Therefore, one needs to study the account of temperature effects in
the ultraslow light phenomenon. Naturally, it is also important to
compare the theoretical results to the experimental
data~\cite{Hau1999N} describing the dependence of the group velocity
of a signal on the temperature of a system.

In the present paper, we generalize the approach developed earlier in
Ref.~\cite{Sly2008PRA} for the uniform (nontrap) systems
on the case of finite (nonzero) temperatures.
This approach is based on the Green-functions
formalism~\cite{Akhiezer1981} and an approximate formulation of the
second-quantization method~\cite{Pel2005JMP}. An object of the
mentioned generalization is a study of the
influence of temperature effects
on the dispersion characteristics of the system and, as a
result, on the propagation properties of a signal in it.

\section{Linear response of a gas in a BEC state
 at finite temperatures: Green functions}

To describe the optical properties of gases consisting of
alkali-metal atoms that are used in the BEC-related experiments, it
is most convenient to use the model of an ideal gas of hydrogenlike
atoms in the stationary state (the limits of this approach are
discussed in Ref.~\cite{Sly2008PRA}). By the term \textquotedblleft stationary
state'' we mean that the atoms can be found only in the states
whose lifetimes are much greater than the relaxation time of the
system (e.g., hyperfine levels of the ground state) and in the
states whose occupation are stimulated by an external
electromagnetic field (e.g., a laser radiation). In the
case of a BEC presence in this gas, the density distribution of atoms in the
quantum state~$\alpha$ by the momentum~$\textbf{p}$ at nonzero
temperatures ($0\leq T\leq T_{\text{c}}$, $T$ is the temperature
in energy units) can be set equilibrium,
therefore, it can be written as follows (see also
Ref.~\cite{Akhiezer1981}):
\begin{equation}\label{eq.t1}                                           
\begin{split}
    \nu_{\alpha}&(\textbf{p})=\nu_{\alpha}\left[1-\left({T}/
    {T_{\text{c}\alpha}}\right)^{3/2}\right]\delta(\textbf{p})
    \\
    &+g_\alpha(2\pi\hbar)^{-3}\{\exp{\left[(\varepsilon_{\alpha}
    (\textbf{p})-\mu_{\alpha})/T\right]}-1\}^{-1},
\end{split}
\end{equation}
where $\nu_{\alpha}$ is the total density of atoms in the $\alpha$
state
\begin{equation*}
    \nu_{\alpha}=\int d\textbf{p}\,\nu_{\alpha}(\textbf{p}),
\end{equation*}
where $T_{\text{c}\alpha}$ is the temperature of the transition of a gas
of atoms in the $\alpha$ state to the BEC phase,
$\delta(\textbf{p})$ is the Dirac delta function, and $g_\alpha=(2F_\alpha+1)$ is the
degeneracy order of the state by the total
momentum~$\textbf{F}_\alpha$ of an atom in this state. In the next
calculations, we assume that an external field is present in the
system, i.e., we set $g_\alpha=1$. Here, also $\hbar$ is the Planck constant,
$\varepsilon_{\alpha}(\textbf{p})=\varepsilon_{\alpha}+p^2/2m$,
where $\varepsilon_{\alpha}$ is the energy of an atom in the $\alpha$ state,
$m$ is the atomic mass ($m=m_{\text{p}}+m_{\text{e}}$;
$m_{\text{p}}$ and $m_{\text{e}}$ are masses of the atomic core
and electron, respectively),
and $\mu_{\alpha}$ is the chemical potential an atom in this
state.

Now we must note the following. It is known that a Bose-Einstein
condensation is a collective effect. The atoms occupying the same quantum
state~$\alpha$ are identical. Because of this fact, one cannot definitely
state what particles participate in the condensation process.
Therefore, the terms \textquotedblleft condensate'' and
\textquotedblleft noncondensate'' particles
do not correspond directly to the selected atoms in the system.
Strictly speaking, the term condensate particles corresponds
only to the fraction of all atoms, which participate in the
formation of the Bose-condensed component. Hence, the number of
noncondensate particles can be found from the difference
between the total number of atoms and number of condensate
particles in the system. By the use of these terms, now we can say
that the first summand in Eq.~(\ref{eq.t1}) corresponds to the
contribution of condensate particles in the distribution function
and the second summand corresponds to the normal component.
Thus in the next calculations, we differentiate contributions from
the different types of particles in the response of the system to the
external perturbation by an electromagnetic field.

It is shown in Ref.~\cite{Sly2008PRA} that the linear response of an
ideal gas of hydrogenlike atoms with BECs to a weak
electromagnetic field can be studied from the first
principles in the framework of the Green-function formalism. There
it was shown that the Fourier transform for the scalar Green
function can be defined by the relation (see also Ref.~\cite{Sly2006CMP}
for details)
\begin{equation}\label{eq.t2}                                              
\begin{split}
    G(\textbf{k},\omega)
    =\dfrac{1}{\mathcal{V}}
    \sum\limits_{\textbf{p}}
    \sum\limits_{\alpha,\beta}
    |\sigma_{\alpha\beta}(\textbf{k})|^2
    \\
    \times
    \dfrac{f_{\alpha}(\textbf{p}-\textbf{k})
    -f_{\beta}(\textbf{p})}
    {\varepsilon_{\alpha}(\textbf{p})-
    \varepsilon_{\beta}(\textbf{p}-\textbf{k})
    +\omega+i\gamma_{\alpha\beta}}.
\end{split}
\end{equation}
Here, by $\textbf{k}$ and $\omega$ we denote the wave vector and the
frequency of the external perturbing field, respectively, and $\gamma_{\alpha\beta}$
denotes the linewidth related to the probability of a spontaneous transition between
the~$\beta$ and $\alpha$ states.

The quantity~$\sigma_{\alpha\beta}(\textbf{k})$ defines the matrix
elements of the charge density of hydrogenlike atoms. This quantity
can be expressed in terms of the wave functions~$\varphi_{\alpha}$ and
$\varphi_{\beta}$
of the atoms in $\alpha$ and $\beta$ states (see also
Refs.~\cite{Sly2008PRA,Pel2005JMP}), respectively,
\begin{eqnarray*}
    &&\sigma_{\alpha\beta}(\textbf{k})
    =e\int d\textbf{y}\,\varphi_{\alpha}^{*}(\textbf{y})
    \varphi_{\beta}(\textbf{y})\nonumber
    \\
    &&\qquad\qquad\times\left[\exp{\left(i\dfrac{m_{\text{p}}}{m}
    \textbf{k}\textbf{y}\right)}
    -\exp{\left(-i\dfrac{m_{\text{e}}}{m}
    \textbf{k}\textbf{y}\right)}\right],
\end{eqnarray*}
where $e$ is the elementary charge.
In particular, in the case when the dipole transition is allowed
between the ${\alpha}$ and ${\beta}$ states, in the linear order over
small term $\textbf{ky}\ll1$, one gets
\begin{equation*}                                                           
    \sigma_{\alpha\beta}(\textbf{k})
    \approx i\textbf{k}\textbf{d}_{\alpha\beta},
    \quad\textbf{d}_{\alpha\beta}=e\int d\textbf{y}\,\textbf{y}
    \varphi_{\alpha}^{*}(\textbf{y})
    \varphi_{\beta}(\textbf{y}),
\end{equation*}
where $\textbf{d}_{\alpha\beta}$ is the matrix element of the atomic
dipole moment. It should be mentioned that in next calculations, we use only the quantities
that are proportional to $|\sigma_{\alpha\beta}(\textbf{k})|^2$. Therefore, for the allowed
dipole transitions, these quantities can be expressed in terms of
the average dipole moment~$d_{\alpha\beta}$,
\begin{equation}\label{eq.t2-2}                                             
    |\sigma_{\alpha\beta}(\textbf{k})|^2
    \approx k^2d^2_{\alpha\beta}/3.
\end{equation}

The quantity~$f_{\alpha}(\textbf{p})$ in Eq.~(\ref{eq.t2})
corresponds to the Bose distribution function of atoms by the
momentum~$\textbf{p}$,
\begin{equation}\label{eq.t3}                                               
    f_{\alpha}(\textbf{p})=
    \{\exp[(\varepsilon_{\alpha}(\textbf{p})
    -\mu_{\alpha})/T]-1\}^{-1}.
\end{equation}
Note that one can use the Bose distribution in the
form~(\ref{eq.t3}) only in the case of the thermal equilibrium in
the system. Thus, in the next calculations, we consider only
the states whose lifetimes are much greater than the system
relaxation time and the states whose existence in the system is
stimulated by an external field. Hence, for a studied
gas (a condensed phase can be formed by atoms with the
energy~$\varepsilon_{\alpha}$), one gets the condition for the
chemical potential~$\mu_{\alpha}$ (see Ref.~\cite{Sly2007LTP} for
details)
\begin{equation*}                                                           
    \mu_{\alpha}(T\leq T_{\text{c}})=\varepsilon_{\alpha}.
\end{equation*}

At temperatures $0\leq T\leq T_{\text{c}}$ in Eq.~(\ref{eq.t2}),
one can substitute the distribution
function~(\ref{eq.t3}) by the density
distribution~(\ref{eq.t1}). To this end, one needs
also use the rule
\begin{equation*}
    \dfrac{1}{\mathcal{V}}
    \sum\limits_{\textbf{p}}f_{\alpha}(\textbf{p})\ldots
    =\int d\textbf{p}~\nu_{\alpha}(\textbf{p})\ldots\,.
\end{equation*}
Next let us note that the relation~(\ref{eq.t2}) is linear over the
distribution function~$f_{\alpha}(\textbf{p})$. Therefore, in
accordance with Eq.~(\ref{eq.t1}), it can be divided into two
summands that define the contributions of the condensate and
noncondensate particles in the system response
\begin{equation}\label{eq.t4}                                               
    G(\textbf{k},\omega)
    =G^{(\text{c})}(\textbf{k},\omega)
    +G^{(\text{n})}(\textbf{k},\omega),
\end{equation}
where the summands $G^{(\text{c})}(\textbf{k},\omega)$ and
$G^{(\text{n})}(\textbf{k},\omega)$ are defined by
\begin{equation}\label{eq.t5}                                               
\begin{split}
    &G^{(\text{c})}(\textbf{k},\omega)
    =\sum\limits_{\alpha,\beta}
    |\sigma_{\alpha\beta}(\textbf{k})|^2
    \dfrac{\nu_{\alpha}(1-t_{\alpha}^{3/2})}
    {\delta\omega_{\alpha\beta}+i\gamma_{\alpha\beta}},
    \\
    &G^{(\text{n})}(\textbf{k},\omega)
    =(2\pi\hbar)^{-3}\sum\limits_{\alpha,\beta}
    |\sigma_{\alpha\beta}(\textbf{k})|^2
    \\
    &\times\int\limits_{0}^{\infty}\dfrac{2\pi p^2dp}
    {\exp{(\varepsilon_{p}/T)}-1}
    \int\limits_{-1}^{1}
    \dfrac{dy}
    {\delta\omega_{\alpha\beta}+pky/m
    +i\gamma_{\alpha\beta}}.
\end{split}
\end{equation}
Here, $t_{\alpha}=T/T_{\text{c}\alpha}$ denotes the relative
temperature of a gas (in this paper, we consider the case
$t_{\alpha}\leq1$), $\delta\omega_{\alpha\beta}
=\omega+\Delta\varepsilon_{\alpha\beta}$ is the frequency
detuning taken in the energy units, $\Delta\varepsilon_{\alpha\beta}=
\varepsilon_{\alpha}-\varepsilon_{\beta}$, $\varepsilon_{p}=p^2/2m$ denotes
the kinetic energy of an atom, and the integration variable $y$ denotes the cosine of the polar
angle~$\theta$, $y\equiv\cos\theta$. Here and below, we neglect of the
recoil energy~$\varepsilon_{\text{r}}=\hbar^2k^2/2m$ that is small enough
to do a significant contribution into the effect
($\varepsilon_{\text{r}}\ll\gamma_{\alpha\beta}$; strictly speaking,
it can be accounted by redefining the
quantity~$\delta\omega_{\alpha\beta}$, i.e., by shifting the
resonant frequency by~$\varepsilon_{\text{r}}$).
For the definiteness, in the next description, the summand
$G^{(\text{c})}(\textbf{k},\omega)$ in Eq.~(\ref{eq.t4}) is called as
the condensate Green function and the summand
$G^{(\text{n})}(\textbf{k},\omega)$ is called as the noncondensate Green
function.

\subsection{Condensate Green function at finite temperatures}

Firstly, we consider the scalar Green function corresponding to
the contribution of condensate particles. For a convenience, let us
study in detail the real and imaginary parts of it. It is shown
below that in the region of transparency the real part makes a main contribution to the refractive
index of a gas and imaginary part makes a main contribution to the absorption rate of light pulses. In accordance with Eq.~(\ref{eq.t5}), the
real and imaginary parts of this function can be written as
\begin{equation}\label{eq.t6}                                               
\begin{split}
    \re G^{(\text{c})}=
    \sum\limits_{\alpha,\beta}
    |\sigma_{\alpha\beta}(\textbf{k})|^2
    \dfrac{\nu_{\alpha}(1-t_{\alpha}^{3/2})\delta\omega_{\alpha\beta}}
    {(\delta\omega_{\alpha\beta})^2+(\gamma_{\alpha\beta})^2},
    \\
    \im G^{(\text{c})}
    =-\sum\limits_{\alpha,\beta}
    |\sigma_{\alpha\beta}(\textbf{k})|^2
    \dfrac{\nu_{\alpha}(1-t_{\alpha}^{3/2})\gamma_{\alpha\beta}}
    {(\delta\omega_{\alpha\beta})^2+(\gamma_{\alpha\beta})^2}.
\end{split}
\end{equation}

\begin{figure}
\includegraphics[%
  width=1.0\linewidth,
    keepaspectratio]{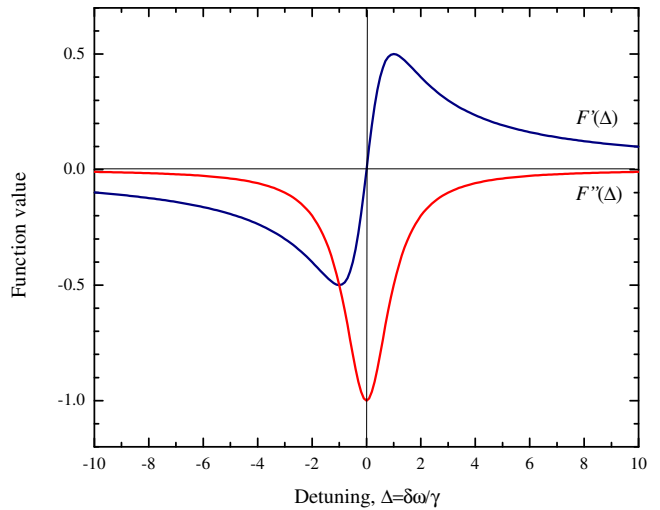} \caption{(Color online) Dependencies of
    the functions $F'$ and $F''$ corresponding to the response of
    the condensate particles on the frequency detuning~$\Delta$.}
  \label{fig.01cond}
\end{figure}

To simplify the description, in the next calculations, we use a model of a
two-level system ($\alpha=1$, $\beta=2$). Below, by the index \textquotedblleft1'' we denote
the set of quantum numbers corresponding to the ground state and by the
index \textquotedblleft2'' we denote the set of quantum numbers corresponding to the
excited state. It is also assumed that the occupation of the exited
state is stimulated by a low-intensity laser pulse. Due to a low
intensity of the pumping field, we can consider that the density of
the atoms in the excited state is small in comparison to the
density of atoms in the ground state (see Ref.~\cite{Sly2008PRA} for
details)
\begin{equation*}
    \nu_{2}\ll\nu_{1}\equiv\nu.
\end{equation*}
According to this relation in Eq.~(\ref{eq.t6}) we neglect of the
summands that are proportional to the density~$\nu_{2}$. Therefore,
one can get
\begin{equation}\label{eq.t7}                                               
\begin{split}
    \re G^{(\text{c})}
    ={|\sigma_{12}(\textbf{k})|^2\nu
    (1-t^{3/2})}
    {\gamma}^{-1}F'(\Delta),
    \\
    \im G^{(\text{c})}
    ={|\sigma_{12}(\textbf{k})|^2\nu
    (1-t^{3/2})}
    {\gamma}^{-1}F''(\Delta),
\end{split}
\end{equation}
where we introduce the quantity $\gamma\equiv\gamma_{12}$ and the
dimensionless functions $F'(x)=x/(x^2+1)$ and $F''(x)=-1/(x^2+1)$.
The dependencies of these functions on the relative frequency
detuning~$\Delta$ ($\Delta=\delta\omega/\gamma$) are shown in
Fig.~\ref{fig.01cond}.

Let us note that the functions $F'$ and $F''$ do not depend on the
temperature. Hence, it is easy to see from Eq.~(\ref{eq.t7}) that
the real and imaginary parts decrease with temperature by a power
law as $t^{3/2}$. Note also that the condensate Green function is
proportional to the density of condensate particles in a gas,
$\nu^{(\text{c})}=\nu(1-t^{3/2})$, and it
equals to zero at the critical temperature, $t=1$.

\subsection{Noncondensate Green function at finite temperatures}

Now we study the scalar Green function~$G^{(\text{n})}$
corresponding to the non-condensate particles (see
Eq.~(\ref{eq.t5})). Taking the real and imaginary parts, for the
two-level system (setting the indexes $\alpha =1$ and $\beta =2$,
see above), we come to the relations
\begin{equation}\label{eq.t8}                                               
\begin{split}
    \re G^{(\text{n})}
    =|\sigma_{12}(\textbf{k})|^2
    \dfrac{m^2T}{4\pi^2\hbar^4k}\,
    I'(\Delta,t),
    \\
    \im G^{(\text{n})}
    =|\sigma_{12}(\textbf{k})|^2
    \dfrac{m^2T}{4\pi^2\hbar^4k}\,
    I''(\Delta,t),
\end{split}
\end{equation}
where we introduce the dimensionless functions ${I}'(\Delta,t)$ and
${I}''(\Delta,t)$ that depend on the frequency of an external field
and on the system parameters (including the temperature of a gas)
\begin{equation}\label{eq.t9}                                               
\begin{split}
    {I}'(\Delta,t)=\int\limits_{0}^{\infty}\dfrac{xdx}
    {e^{x^2}-1}\ln
    \left|{\dfrac{1+[\Delta+x(\kappa\sqrt{t})]^2}
    {1+[\Delta-x(\kappa\sqrt{t})]^2}}\right|,
    \\
    {I}''(\Delta,t)=2\int\limits_{0}^{\infty}\dfrac{xdx}
    {e^{x^2}-1}\left\{\arctan\left[{\Delta-x(\kappa\sqrt{t})}
    \right]\right.
    \\
    \left.-\arctan\left[{\Delta+x(\kappa\sqrt{t})}
    \right]\right\}.
\end{split}
\end{equation}
Here, analogously to the formulas for the condensate particles (see
Eq.~(\ref{eq.t7})), $\Delta=\delta\omega/\gamma$ is the relative
frequency detuning, $t=T/T_{\text{c}}$ is the relative temperature,
\begin{equation}\label{eq.t10}                                              
    T_{\text{c}}=2\pi[\zeta(3/2)]^{-2/3}
    \hbar^2\nu^{2/3}m^{-1}
\end{equation}
is the critical temperature of the transition to the BEC phase,
and $\zeta(x)$ is the Riemann zeta function. In Eq.~(\ref{eq.t9}), we also
introduce the dimensionless parameter~$\kappa$,
\begin{equation}\label{eq.t11}                                              
    \kappa=\dfrac{\hbar k}
    {\gamma}\sqrt{\dfrac{2T_{\text{c}}}
    {m}}\sim\dfrac{\hbar k}
    {\gamma m}\hbar\nu^{1/3}
\end{equation}
that depends on the physical properties of a gas and the wave number~$k$ of the external electromagnetic field.
It is shown below that the value of this parameter defines the
characteristics of the normal component response to an external perturbation.

\begin{figure}
\includegraphics[%
  width=1.0\linewidth,
  keepaspectratio]{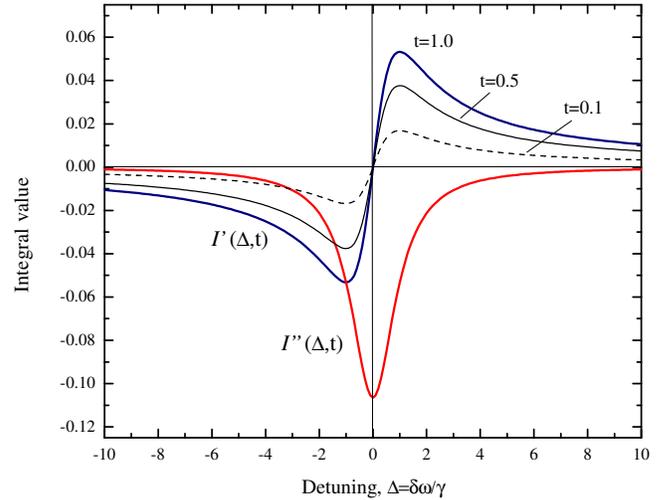} \caption{(Color online) Dependencies of
  the functions ${I}'(\Delta,t)$ and ${I}''(\Delta,t)$ on the
  frequency detuning at $T=T_\text{c}$ (bold lines). By thin lines,
  the dependencies of the function ${I}'$ at temperatures
  $T/T_\text{c}=0.5$ and 0.1 are shown ($\kappa=0.025$).}
  \label{fig.02int}
\end{figure}

Let us estimate the value of the parameter~$\kappa$ for the resonant
radiation that corresponds to the sodium $D_2$ line. The laser pulses
tuned to the components of this line were used in
Ref.~\cite{Hau1999N}. Taking $m=3.82\times10^{-23}$~g,
$\gamma=8.1\times10^{-21}$~erg, $k=1.07\times10^{5}$~cm$^{-1}$, and
$T_c=435$~nK, we get $\kappa\approx0.025$. Using this parameter value it is easy to get the dependencies for the integrals ${I}'(\Delta,t)$ and
${I}''(\Delta,t)$ that are shown in Fig.~\ref{fig.02int}.
There, one can study also a behavior of the integral~$I'$ with a decrease
of the temperature to the values $0.5T_c$ and $0.1T_c$, respectively
(note that the integral~$I''$ has an analogous temperature dependence).

From Fig.~\ref{fig.02int}, one can see that the dependencies of the
functions~$I'$ and $I''$ are similar to the dependencies of the
condensate functions $F'$ and $F''$ (cf. Fig.~\ref{fig.01cond}). The
reasons for this similarity become evident below, but now it is easy
to see that the values of the functions~$I'$ and~$I''$ strongly
depend on the gas temperature. In particular, from Eq.~(\ref{eq.t8})
and Fig.~\ref{fig.02int}, we can conclude that the contribution of
the non-condensate Green functions in the total Green
function~(\ref{eq.t4}) increases with temperature. Let us study
this effect in detail in the next subsection.

\subsection{Dependence of the scalar Green function on the temperature}

Let us note that the functions~$I'$ and $I''$ characterizing the
contribution of the non-condensate particles to the total response
of a gas in the case $\kappa\ll1$ can be simplified at arbitrary
values of the detuning~$\Delta$. Really, due to the strongly decreasing function~$\exp(-x^2)$ in the integrand, the main contribution to
the integral give small values of the integration variable~$x$, $x\lesssim1$.
This fact allows us to expand the integrands into series
over~$(x\kappa\sqrt{t})\ll1$. As a result, using Eqs.~(\ref{eq.t7})
and (\ref{eq.t9}), accurate within quadratic summands, we get
\begin{equation}\label{eq.t12}                                        
\begin{split}
    &{I}'(\Delta,t)\approx4\kappa\sqrt{t}\,
    {F}'(\Delta)\left\{\int\limits_{0}^{\infty}\dfrac{x^2dx}
    {e^{x^2}-1}\right.
    \\
    &~~\left.+(\kappa\sqrt{t})^2
    \left[\dfrac{4}{3}F'^2(\Delta)+F''(\Delta)\right]
    \int\limits_{0}^{\infty}\dfrac{x^4dx}{e^{x^2}-1}\right\},
    \\
    &{I}''(\Delta,t)\approx4\kappa\sqrt{t}\,
    {F}''(\Delta)\left\{\int\limits_{0}^{\infty}\dfrac{x^2dx}
    {e^{x^2}-1}\right.
    \\
    &~~\left.+(\kappa\sqrt{t})^2
    \left[F'^2(\Delta)-\dfrac{1}{3}F''^2(\Delta)\right]
    \int\limits_{0}^{\infty}\dfrac{x^4dx}{e^{x^2}-1}\right\}.
\end{split}
\end{equation}

According to the relations
\begin{equation*}                                                   
    \int\limits_{0}^{\infty}\dfrac{x^2dx}
    {e^{x^2}-1}=\dfrac{\sqrt{\pi}}{4}\zeta(3/2),\quad
    \int\limits_{0}^{\infty}\dfrac{x^4dx}
    {e^{x^2}-1}=\dfrac{3\sqrt{\pi}}{8}\zeta(5/2),
\end{equation*}
the Green functions~(\ref{eq.t8}) with account of
Eqs.~(\ref{eq.t10})--(\ref{eq.t12}) can be written as
\begin{equation}\label{eq.t13}                                    
\begin{split}
    &\re G^{(\text{n})}
    \approx|\sigma_{12}|^2\nu
    t^{3/2}{\gamma}^{-1}F'(\Delta)[1+\kappa^2tS'(\Delta)],
    \\
    &\im G^{(\text{n})}
    \approx|\sigma_{12}|^2\nu
    t^{3/2}{\gamma}^{-1}F''(\Delta)
    [1+\kappa^2tS''(\Delta)],
\end{split}
\end{equation}
where we introduced the functions
\begin{equation*}
\begin{split}
    &S'(\Delta)=2
    \dfrac{\zeta(5/2)}{\zeta(3/2)}
    \left[F'^2(\Delta)+\dfrac{3}{4}
    F''(\Delta)\right],
    \\
    &S''(\Delta)=\dfrac{3}{2}
    \dfrac{\zeta(5/2)}{\zeta(3/2)}
    \left[F'^2(\Delta)-\dfrac{1}{3}F''^2(\Delta)\right].
\end{split}
\end{equation*}
Note that Eq.~(\ref{eq.t13}) explains also the similarity in the
behavior of the functions~$F'$,~$F''$ and $I'$,~$I''$ at
$(\kappa\sqrt{t})\ll1$ that is mentioned above (see
Figs.~\ref{fig.01cond} and \ref{fig.02int}).

It is easy to see that the \textquotedblleft total'' Green function
in the region from the absolute zero to $T_c$ weakly depends on the
temperature. This dependence appears only in the quadratic terms
over~$(\kappa\sqrt{t})\ll1$. Really, by the use of
Eqs.~(\ref{eq.t4}), (\ref{eq.t7}), and (\ref{eq.t13}), one can get
\begin{equation}\label{eq.t14}                                    
\begin{split}
    &\re G\approx
    |\sigma_{12}|^2\nu
    {\gamma}^{-1}F'(\Delta)
    [1+\kappa^2tS'(\Delta)],
    \\
    &\im G\approx
    |\sigma_{12}|^2\nu
    {\gamma}^{-1}F''(\Delta)
    [1+\kappa^2tS''(\Delta)].
\end{split}
\end{equation}

\begin{figure}
\includegraphics[%
  width=1.0\linewidth,
  keepaspectratio]{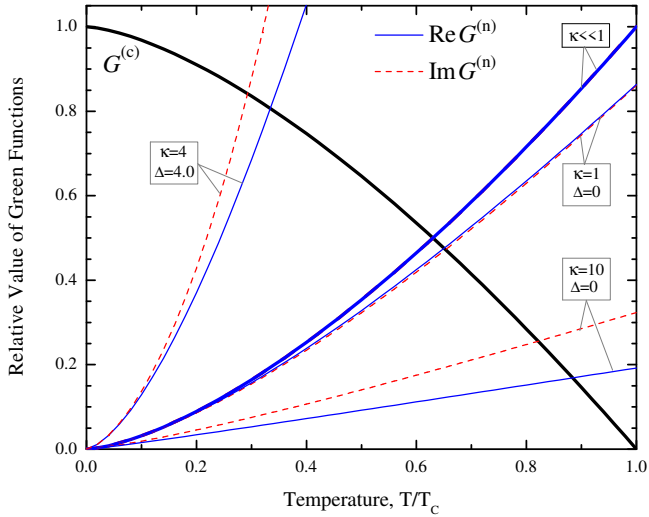}
  \caption{(Color online) Relative dependencies
  of the condensate and noncondensate Green functions on the
  temperature. All functions are normalized to a value of the
  condensate Green function at $t=0$ and corresponding
  detuning~$\Delta$.}
  \label{fig.03relval}
\end{figure}

\begin{figure}
\includegraphics[%
  width=1.0\linewidth,
  keepaspectratio]{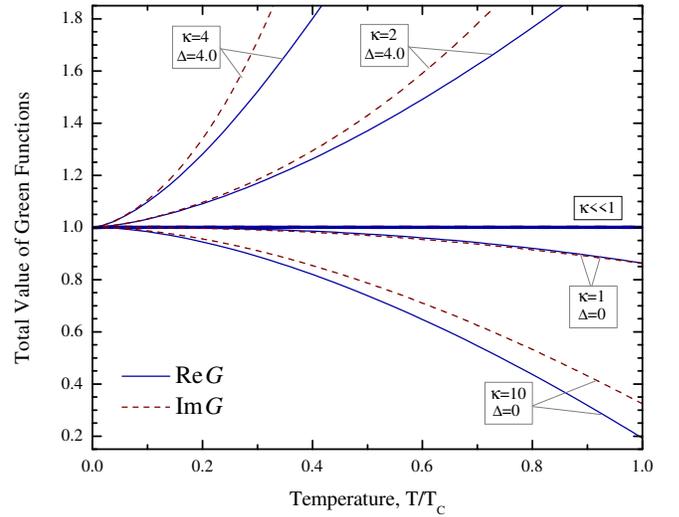} \caption{(Color online) Relative dependencies
  of the total Green functions on the
  temperature. All functions are normalized to a value of the
  condensate Green function at $t=0$ and corresponding
  detuning~$\Delta$.}
  \label{fig.04totval}
\end{figure}

But in the case $(\kappa\sqrt{t})\gtrsim1$, the
expansions~(\ref{eq.t12}) become incorrect and analytical formulas
for the functions~$I'$ and $I''$ cannot be found. Therefore, in this
case, we use the numerical calculations at the definite values of the
detuning~$\Delta$ and parameter~$\kappa$ (see
Figs.~\ref{fig.03relval} and \ref{fig.04totval}).

In Fig.~\ref{fig.03relval}, one can see the dependencies of the
condensate and noncondensate Green functions on the temperature.
The real parts of these functions  are normalized to a value of the
real part of the condensate Green function at $t=0$ and
corresponding detuning~$\Delta$. Analogously, the imaginary parts of
these functions  are normalized to a value of the imaginary part of
the condensate Green function at $t=0$ and corresponding
detuning~$\Delta$. In other words, in Fig.~\ref{fig.03relval}, the
following dependencies are shown:
\begin{equation*}
\begin{split}
    \re G^{(\text{c},\text{n})}(t,\Delta)/
    \re G^{(\text{c})}(t=0,\Delta),
    \\
    \im G^{(\text{c},\text{n})}(t,\Delta)/
    \im G^{(\text{c})}(t=0,\Delta).
\end{split}
\end{equation*}
One can see that due to this normalization, the inequality
$\im G(t,\Delta)/\im G^{(\text{c})}(0,\Delta)>0$ takes
place. But note that the imaginary part of all Green functions is
negative due to the  function~$F''(\Delta)$ (see Eqs.~(\ref{eq.t7})
and (\ref{eq.t13})).

Analyzing the dependencies shown in Fig.~\ref{fig.03relval} one can
conclude that the real and imaginary parts of the condensate Green
function demonstrate a similar behavior (drop-down bold curve).
This dependence agrees with
the obtained relations. But at the same time, the relative values of
the real and imaginary parts of the noncondensate Green functions
strongly depend on the parameter~$\kappa$ and detuning~$\Delta$. One
can conclude that at $\kappa\lesssim 1$, by the use of the mentioned
normalization, the real and imaginary parts are practically coincide.
In the opposite case, $\kappa >1$, as one can see in
Fig.~\ref{fig.03relval}, the real and imaginary parts demonstrate different
dependencies. This fact must have a strong impact
on the dependence of the total Green function (the sum of the
condensate and non-condensate Green functions) on the temperature.

In Fig.~\ref{fig.04totval},
the dependencies of the total Green function on the temperature are
shown at different values of the parameters~$\Delta$ and $\kappa$.
Here, it is important to pay attention to the central curve that is
practically horizontal (the case $\kappa\ll 1$). This curve demonstrates a weak dependence of
the total Green function (including its real and imaginary parts) on
the temperature. Note that one can come to the same conclusion by analyzing
Eq.~(\ref{eq.t14}). Thus at $\kappa\ll 1$, the influence of the
temperature effects in the response of a BEC to an external field is
insignificant. The problem on obtaining the response of this system
can be solved with a good accuracy by setting the temperature of a
gas equal to zero (exactly this approximation was used in
Refs.~\cite{Sly2008PRA,Sly2006CMP,Sly2008JLTP,Sly2009PLA}).
In other cases (at $\kappa >1$), as one can
see from Fig.~\ref{fig.04totval}, the influence of the temperature
effects can be rather significant.

Evidently, the effects studied in this section in some way
(significantly or not) can influence on the dispersion
properties of a condensed gas. Let us recall that the resonant peculiarities
of the dispersion characteristics play an essential role in the ultraslow-light
phenomenon in a BEC of alkali-metal atoms.

\section{Dispersion characteristics of a gas in a BEC state at finite
            temperatures}

It is known that in the framework of the linear approach,
the permittivity~$\epsilon(\textbf{k},\omega)$ of a
gas can be expressed in terms of the scalar Green function (see in
this case Refs.~\cite{Akhiezer1981,Sly2006CMP}) as
\begin{equation*}\label{eq.t15}                                           
    \epsilon^{-1}(\textbf{k},\omega)
    =1+\dfrac{4\pi}{k^2}G(\textbf{k},\omega).
\end{equation*}
Therefore, the real and imaginary parts of the
permittivity~$\epsilon(\textbf{k},\omega)$,
\begin{equation*}                                                         
    \epsilon(\textbf{k},\omega)=\epsilon'(\textbf{k},\omega)
    +i\epsilon''(\textbf{k},\omega)
\end{equation*}
can be written as
\begin{equation}\label{eq.t16}                                            
\begin{split}
    \epsilon'(\textbf{k},\omega)=\dfrac{1+4\pi k^{-2}\re G}
    {(1+4\pi k^{-2}\re G)^2+(4\pi k^{-2}\im G)^2},
    \\
    \epsilon''(\textbf{k},\omega)=\dfrac{-4\pi k^{-2}\im G}
    {(1+4\pi k^{-2}\re G)^2+(4\pi k^{-2}\im G)^2}.
\end{split}
\end{equation}
At the same time, the refractive index and damping factor can be
written in terms of the real and imaginary parts of the
permittivity,
\begin{equation}\label{eq.t17}                                             
\begin{split}
    n'(\textbf{k},\omega)=\dfrac{1}{\sqrt{2}}
    \sqrt{\sqrt{\epsilon'^2+\epsilon''^2}+\epsilon'},
    \\
    n''(\textbf{k},\omega)=\dfrac{1}{\sqrt{2}}
    \sqrt{\sqrt{\epsilon'^2+\epsilon''^2}-\epsilon'}.
\end{split}
\end{equation}
Hence, basing on the derived equations for the scalar Green
functions, one can study the propagation properties of weak
electromagnetic pulses through a BEC at finite (nonzero)
temperatures.
As the derived analytical expressions for the refractive index and
damping factor in the general case may have rather lengthy form (see
Eq.~(\ref{eq.t5})), let us consider some particular cases.

\subsection{Temperature effects in two-level systems}

Firstly, let us consider a case when the frequency of an external
field is close to the energy spacing between two definite quantum
states of atoms. Thus, using the most common method (see, e.g.,
Ref.~\cite{Allen1987}), in calculations we can consider only the
resonant terms corresponding to this transition. In other words, we
can consider atoms as a two-level system. Therefore, by the use of
Eqs.~(\ref{eq.t4}), (\ref{eq.t7}), (\ref{eq.t8}), and
(\ref{eq.t11}), the real and imaginary parts of the permittivity
(see Eq.~(\ref{eq.t16})) take the form
\begin{equation}\label{eq.t18}                                         
\begin{split}
    \epsilon'(\textbf{k},\omega;t)=\dfrac{1+aG_1}
    {(1+aG_1)^2+(aG_2)^2},
    \\
    \epsilon''(\textbf{k},\omega;t)=\dfrac{-aG_2}
    {(1+aG_1)^2+(aG_2)^2},
\end{split}
\end{equation}
where
\begin{equation}\label{eq.t19}                                         
\begin{split}
    &G_1=(1-t^{3/2})F'(\Delta)+btI'(\Delta,t;b),
    \\
    &G_2=(1-t^{3/2})F''(\Delta)+btI''(\Delta,t;b),
\end{split}
\end{equation}
\begin{equation}\label{eq.t20}                                         
    a=4\pi\dfrac{|\sigma_{12}(\textbf{k})|^2\nu}{k^2\gamma},
    \quad
    b=[\kappa\sqrt{\pi}\zeta(3/2)]^{-1}.
\end{equation}

\begin{figure}
\includegraphics[%
    width=1.0\linewidth,
    keepaspectratio]{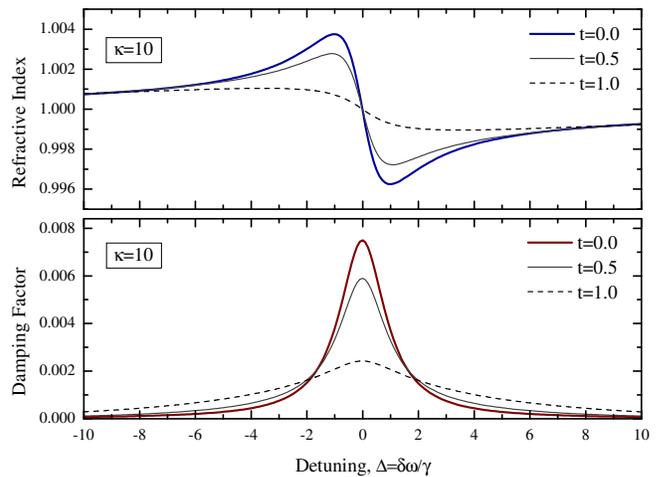} \caption{(Color online)
    Dispersion dependencies of the two-level system in the BEC
    state. The used value for the parameter $a=0.015$.}
    \label{fig.04disp}
\end{figure}

In particular, for the condensed sodium vapor with the density
$\nu=1.44\times10^{12}$~cm$^{-3}$ that interacts with the resonant
radiation corresponding to the $D_2$ line, in accordance with
Eqs.~(\ref{eq.t2-2}) and (\ref{eq.t20}), we get
$a\approx4.19d^2\nu/\gamma$. Next, considering $d^2\approx
S_{FF'}(3.52er_0)^2$ ($r_0$ is the Bohr radius, $e$ is the
elementary charge, and $S_{FF'}$ is the relative strength of
the dipole-allowed transition $F\rightarrow F'$ \cite{Steck2000}),
$F=2$, $F'=2$, and $S_{22}=1/4$, we come to $a\approx0.015$ and
$b\approx20.38$. For the two-level system with these parameters, the
characteristic dependencies are shown in Fig.~\ref{fig.04disp}
(straight bold lines on the upper and lower graphs;
there also the dependencies for the case $\kappa=10$ are shown).
Let us note that these dependencies (bold lines) also correspond
to the case $t=0$ independently of the parameter~$\kappa$ value
(see also Eq.~(\ref{eq.t14})).

\begin{figure}
\includegraphics[%
  width=1.0\linewidth,
  keepaspectratio]{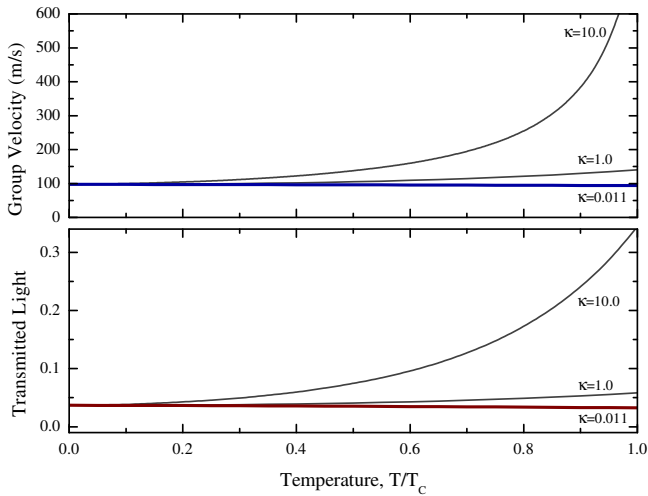} \caption{(Color online) Dependencies of the group
  velocity and intensity of the transmitted light on the temperature
  for the two-level system with different values of the
  parameter~$\kappa$.}
  \label{fig.05temp}
\end{figure}

Now we use the expressions that define the values of the group
velocity and intensity of the transmitted light
\begin{equation}\label{eq.t21}                                         
\begin{split}
    &v_{\text{g}}(t)=c
    \{n'(t)+\omega[\partial n'(t)/\partial \omega]\}^{-1},
    \\
    &I(t)=I_{0}\exp{[-n''(t)kL]},
\end{split}
\end{equation}
where $L$ is the characteristic size of the atomic cloud (in
calculations we set  $L=0.004$~cm). For the mentioned physical
characteristics, one can obtain the dependencies corresponding to the
signals tuned up exactly to the resonant frequency ($\Delta=0$). The
corresponding curves are shown in Fig.~\ref{fig.05temp}.
As one can conclude from Fig.~\ref{fig.05temp}, the propagation
velocity of the light pulse at the temperatures lower than the critical
temperature in the case $\kappa\ll1$ (as it is expected) practically
does not depend on the temperature.

Note that inequality~$\kappa\ll1$ usually takes place for dilute
gases of alkali-metal atoms that interact with a laser field
($\kappa\approx0.025$ for the parameters of the
experiment~\cite{Hau1999N}). Probably, this inequality takes place
for the most BEC-related experiments at the present moment. Really,
as it comes from the definition~(\ref{eq.t11}) of the
parameter~$\kappa$, to increase its value in one order of magnitude,
the density of atoms in a condensate must be increased by 3
orders of magnitude. This requirement results in the instability of
the BEC phase due to an increase of the number of three-body
collisions (see Ref.~\cite{Cor1999ISP}). Also it is easy to
see that the use of the high-frequency radiation (with the large
wave number~$k$) cannot result to the considerable increase of the
parameter~$\kappa$. It comes from the fact that the high-frequency
transitions correspond to the high levels of the atomic spectrum.
But the linewidth of these levels is much greater than the linewidth
corresponding to the lower states. Hence, probably, the only way to
a significant increase of the $\kappa$ value is to use the states
with the smaller linewidth~$\gamma$ (see Eq.~(\ref{eq.t11})). It
means that to increase the mentioned value, one needs to select the
levels whose probability of a spontaneous transition is much less
than was used in the above calculations.

Therefore, the cases $\kappa\gtrsim1$ that are shown in
Fig.~\ref{fig.05temp} correspond to the \textquotedblleft
long-living'' levels. One may relate these states to the levels with
the forbidden dipole transitions (e.g., hyperfine levels of the
ground state, see in this case Ref.~\cite{Sly2008JLTP}) or the
sublevels, whose relative transition intensity is rather low. One
can conclude that in this case, the temperature effects must have a
strong impact on the ultraslow light phenomenon in a BEC.

\subsection{Temperature effects in three-level systems}

Note that two-level systems from the standpoint of the experiments
dealing with the ultraslow-light phenomenon in a BEC can be
inconvenient. As it is known, this phenomenon is realized mostly in
three-level systems. These systems have some characteristic
advantages: the absorption rate of the signal can be decreased by a
special tuning of the laser frequency, one can get positive time
delays of the pulses (positive sign of the group velocity), and also
one can use the magnetic field to control the group velocity of the
ultraslow pulses~\cite{Sly2009PLA} or use an additional coupling
laser to provide the electromagnetically induced
transparency~\cite{Har1997PT}.

\begin{figure}
\includegraphics[width=1.0\linewidth,keepaspectratio]{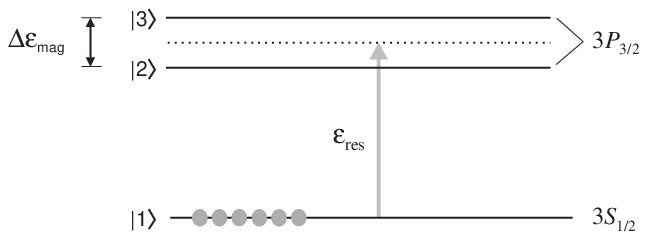}\vspace{3mm}\\
\includegraphics[width=1.0\linewidth, keepaspectratio]{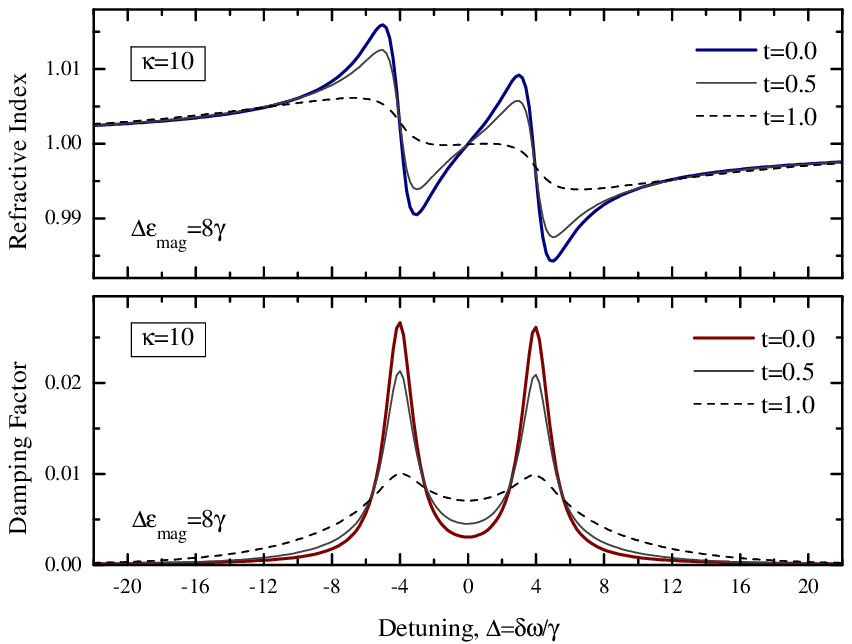}
    \caption{(Color online) Scheme and dispersion dependencies
    of the three-level system in a BEC
    state. The used value for the parameter $a=0.052$.}
    \label{fig.07disp2}
\end{figure}

Therefore, in the framework of the developed approach let us study
the temperature dependencies of the group velocity and intensity of
the transmitted light in a three-level system. To this end, let us
consider the system that is schematically illustrated in
Fig.~\ref{fig.07disp2} (upper part).

Note that in the framework of the developed approach, we do not
account quantum interferences in the system. Therefore, in this case,
the derived above expressions change insignificantly.
In particular, due to the additive
contribution of all quantum states to the Green functions (see
Eq.~(\ref{eq.t5})), the functions $G_1$ and $G_2$ characterizing the
response of all particles in the system can be written as follows
(cf. Eq.~(\ref{eq.t19})):
\begin{equation}\label{eq.3.9}                                                 
\begin{split}
    G_1=&(1-t^{3/2})[F'(\Delta+\Omega/2)+F'(\Delta-\Omega/2)]
    \\&+bt[I'(\Delta+\Omega/2,t;b)+I'(\Delta-\Omega/2,t;b)],
    \\
    G_2=&(1-t^{3/2})[F''(\Delta+\Omega/2)+F''(\Delta-\Omega/2)]
    \\&+bt[I''(\Delta+\Omega/2,t;b)+I''(\Delta-\Omega/2,t;b)],
\end{split}
\end{equation}
where $\Omega=\Delta\varepsilon_{\text{mag}}/\gamma$ and
$\Delta\varepsilon_{\text{mag}}$ is the energy spacing between the
excited states. Here and below, we consider that the splitted levels
belong to the same multiplet, i.e.,
$|\sigma_{12}|^2/\gamma_{12}=|\sigma_{13}|^2/\gamma_{13}$. Hence, in
accordance with Eq.~(\ref{eq.t20}), we can set $a_{12}=a_{13}=a$.
Thus the formulas~(\ref{eq.t16}), (\ref{eq.t17}), and (\ref{eq.t21})
that define the dependence of the dispersion characteristics on the
temperature have the same form.

To get the numerical estimates, let us consider the condensed gas of
sodium atoms with the density $\nu=5\times10^{12}$~cm$^{-3}$. For
this gas, we get $a=0.052$, $\kappa=0.016$, and $b=13.46$. Also we
consider that there are two nearby excited levels in the system
(states $|2\rangle$ and $|3\rangle$ in Fig.~\ref{fig.07disp2}). We
also set the spacing between them several times larger than the
linewidth~$\gamma$, $\Delta\varepsilon_{\text{mag}}=8\gamma$.
For this system, one can get the graphs characterizing the dispersion
characteristics of the system that are shown in Fig.~\ref{fig.07disp2}
(bold curves). Note that in this case ($\kappa\ll1$), the functions $n'(\omega)$
and $n''(\omega)$ practically do not depend on the temperature. Let us emphasize that we
regard the system as a superposition of two two-level systems (Autler-Townes treatment). This
is the reason that the absorption rate does not go to zero at $\Delta=0$
(Fig.~\ref{fig.07disp2}, bottom) as in the electromagnetically-induced
transparency (EIT) regime \cite{Hau1999N,Har1997PT}.

Within the developed approach, one can also study the dependencies in the case when the
temperature effects have a significant influence on the system response
(as it is mentioned above, this corresponds to the case $\kappa>1$). In particular,
taking $\kappa=10$ and the same values of the other parameters, we get the
dispersion curves at different temperatures that are shown in Fig.~\ref{fig.07disp2}.
It should be noted that in this case, the steepness of the slope of the
refractive index decreases with the temperature. At the same time, the absorption
in the central region increases with the temperature. This behavior can be explained
by the fact that the density distribution of the non-condensate particles by the momentum~\textbf{p} in the case
$\kappa>1$ is not so \textquotedblleft sharp''
as for the condensate particles (see Eq.(\ref{eq.t1})). As a result, the absorption rate of the pulses tuned up to the resonance ($\Delta<1$) decreases and the absorption rate of the pulses detuned from the resonance ($\Delta>1$) increases (see Figs.~\ref{fig.04totval}, \ref{fig.04disp}, and \ref{fig.07disp2}) with the temperature.

\begin{figure}
\includegraphics[%
  width=1.0\linewidth,
  keepaspectratio]{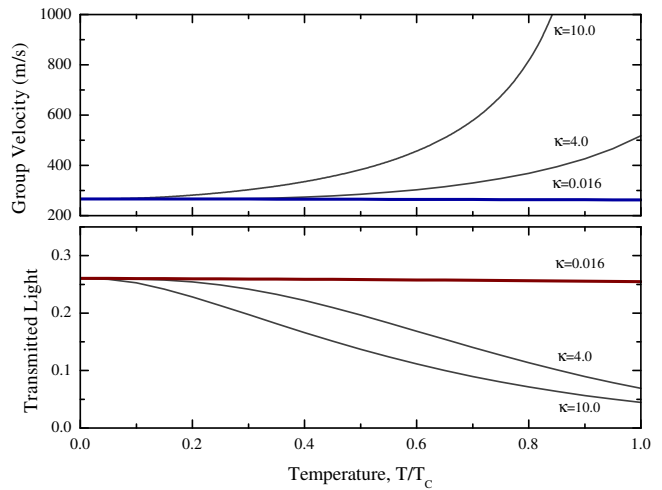} \caption{(Color online) Dependencies of the group
  velocity and intensity of the transmitted light on the temperature
  for the three-level system with different values of the
  parameter~$\kappa$.}
  \label{fig.08temp2}
\end{figure}
The mentioned behavior of the main macroscopic parameters with an
account of Eq.~(\ref{eq.t21}) results in the dependencies
that are shown in Fig.~\ref{fig.08temp2}.
As one can see from the graphs, the velocity of a light pulse at the
temperature lower than the critical temperature in the case
$\kappa\ll 1$ (analogously to the two-level case) is practically
constant. It corresponds to the fact that the refractive index profile and absorption
rate in this case changes insignificantly. Let us emphasize
that in this case for the calculations we used the same parameters
of a gas with a BEC as were realized in Ref.~\cite{Hau1999N}.

At $\kappa>1$, the situation differs from the
two-level case. Let us note that here one can see that both the
group velocity and absorption rate of the pulse increase with the
temperature. It can be explained by the fact that for the detuned pulses the
steepness of the slope of the refractive index decreases while the absorption
increases with the temperature (see Fig.~\ref{fig.07disp2}).
Thus the system becomes not so convenient for the realization of the
ultraslow light phenomenon as in the zero-temperature limit.

\section{Conclusion}

In this paper, we studied some optical properties of dilute gases in
the BEC state at finite temperatures. The analysis is based on the
Green-function formalism. We found analytical expressions for the
scalar Green functions that characterize a linear response of a gas
to an external electromagnetic field (laser). We studied the
characteristic dependencies of these functions on the temperature
and frequency detuning. We made a comparison of the contribution of
the condensate and non-condensate particles into the effect. The
ultraslow-light phenomenon in a BEC is studied both on the examples
of two-level and three-level systems.

In particular, it is shown that for a light pulse, which is tuned up
close to the dipole-allowed transitions, the dispersion
characteristics of a gas of alkali-metal atoms weakly depend on the
temperature in the region from the absolute zero to the critical
temperature. Therefore, the group velocity of the pulses in this
system weakly depends on the temperature. This fact
allowed us to conclude that the results of the previous works, where
the authors studied the response of a gas in the limit of zero
temperatures, $T\rightarrow0$, are correct and can be used also in
the region $0\leq T\leq T_{\text{c}}$.

It is significant to note that in the experiment~\cite{Hau1999N},
data relating to the dependence of the group velocity of light
pulses on the temperature of an ultracold gas of sodium atoms were
obtained. From these results, one can conclude that the group velocity
weakly depends on the temperature in the region $0<T<T_{\text{c}}$.
In the present paper, in the case of a three-level system, we use the
same parameters of a gas as in Ref.~\cite{Hau1999N} and we come to
the same conclusion in our research. But the statement that the
results of our theory are verified by the mentioned experiment is not
quite correct. It should be noted that in Ref.~\cite{Hau1999N}, the
effect of the electromagnetically induced transparency was used. But
in the framework of the introduced approach that is based on the
Green-functions formalism, we cannot account for quantum
interference effects in the system (see in this case
Ref.~\cite{Sly2008PRA}). Probably, the similarity of the results may
be a sign that the quantum interference effects in the mentioned
experiment have a weak influence on the dependence of the group
velocity on the temperature of a gas with a BEC. But, obviously, this
statement requires an additional experimental verification.

We also studied the cases when the temperature effects
can have a strong impact on the dispersion characteristics of gases
with a BEC. In our opinion, this situation may be realized when the
frequency of an external field is tuned close to the transitions between
\textquotedblleft long-living'' states. These states correspond to
the excited levels with the forbidden dipole transitions (upper
hyperfine levels of the ground state of alkali-metal atoms) or the
levels with a low relative intensity of the transition. In this
case, as one can conclude from the results of the paper, the
parameters of slowing impair with the temperature increase. Thus, the
achievement of the lower temperatures of a gas is more
necessary in this case.

\acknowledgments{This work is partly supported by the National Fund
of Fundamental Research of Ukraine Grant No. 25.2/102 and
the National Academy of Sciences of Ukraine Grant No.
55/51–2009.}

\end{document}